\documentstyle[aaspp4,11pt,psfig]{article}   

\newcommand {\be} {\begin{equation}}
\newcommand {\ee} {\end{equation}}

\def\chisi{\chi^s_i}
\def\chis1{\chi^s_1}
\def\chis2{\chi^s_2}
\def\chiai{\chi^a_i}
\def\chia1{\chi^a_1}
\def\chia2{\chi^a_2}
\def\scatt{\chi^s_{ij}(\mu,\mu^{\prime})}
\def\ascatt{\chi^s_{ji}(\mu,\mu^{\prime})}
\def\mup{\mu^{\prime}}
\def\ii{I_E^i}
\def\iip{I_E^{i+}}
\def\iim{I_E^{i-}}
\def\ij{I_E^j}
\def\dii{\partial I_E^i}
\def\tes{\tau_{\rm T}}
\def\dtes{\partial \tau_{\rm T}}
\def\taui{\tau_{i}}
\def\dtaui{\partial \tau_{i}}
\def\ui{u^{i}}
\def\vi{v^{i}}
\def\si{S^{i}}
\def\sig{\sigma_T}
\def\bnu{B_{E}(T)}

\def\ns{neutron star}
\def\nss{neutron stars}
\def\cfp{\chi_F(\tes^\prime)}
\def\cfpp{\chi_F^\prime(\tes^\prime)}

\def\cp{\chi_P}
\def\cj{\chi_J}
\def\dh{\Delta H}
\def\dt{\Delta T}
\def\dtp{\Delta T(\tes^\prime)}
\def\dtesp{\rm{d}\tes^\prime}
\def\tesp{\tes^\prime}

\begin{document}

\title{Surface Emission Properties of Strongly Magnetic Neutron Stars}

\author{Feryal \"Ozel}
\affil{Physics Department, Harvard University and 
Harvard-Smithsonian Center for Astrophysics \\
 60 Garden St., Cambridge, MA 02138;\\ fozel@cfa.harvard.edu}

 \begin{abstract} 

 We construct radiative equilibrium models for strongly magnetized
 ($B\gtrsim 10^{13}$~G) neutron-star atmospheres taking into account
 magnetic free-free absorption and scattering processes computed for
 two polarization modes. We include the effects of vacuum polarization
 in our calculations. We present temperature profiles and the angle-,
 photon energy-, and polarization-dependent emerging intensity for a
 range of magnetic field strengths and effective temperatures of the
 atmospheres. We find that for $B\lesssim 10^{14}$~G, the emerging
 spectra are bluer than the blackbody corresponding to the effective
 temperature, $T_{\rm eff}$, with modified Planckian shapes due to the
 photon-energy dependence of the magnetic opacities. However, vacuum
 polarization resonance significantly modifies the spectra for $B\sim
 10^{15}$~G, giving rise to power-law tails at high photon energies.
 The angle-dependence (beaming) of the emerging intensity has two
 maxima: a narrow (pencil) peak at small angles ($\lesssim 5^\circ$)
 with respect to the normal and a broad maximum (fan beam) at
 intermediate angles ($\sim 20-60^\circ$). The relative importance and
 the opening angle of the radial beam decreases strongly with
 increasing magnetic field strength and decreasing photon energy. We
 finally compute a $T_{\rm eff}-T_{\rm c}$ relation for our models,
 where $T_{\rm c}$ is the local color temperature of the spectrum
 emerging from the \ns\ surface, and find that $T_{\rm c}/T_{\rm eff}$
 ranges between $1.1-1.8$. We discuss the implications of our results
 for various thermally emitting \ns\ models.
 
 \end{abstract}


\section{INTRODUCTION}

The magnetized surface layers of young neutron stars strongly affect
their observable characteristics, as well as many physical processes
that take place in their interiors. Understanding the properties of
these layers becomes most important when the questions under study are
related to the transport of thermal energy from an internal source to
the surface of \nss, as the magnetized envelopes determine the global
thermodynamic properties of the \ns\ in addition to generating the
thermal emission spectra. Strong magnetic fields ($B \gtrsim
10^{12}$~G) significantly modify the plasma properties and the
transport of radiation through the \ns\ atmospheres. Thermal emission
from \nss\ with such magnetic fields has been observed as optical and
soft X-ray emission from young radio pulsars (e.g., \"Ogelman 1995)
and has also recently been suggested as the origin of the X-ray
emission from two new classes of X-ray pulsars.

Neutron stars with ultrastrong magnetic fields ($B \gtrsim 10^{14}$
G), or magnetars (Thompson \& Duncan 1993), currently provide a very
promising explanation for these two classes of sources, the Anomalous
X-ray Pulsars (AXPs; Mereghetti \& Stella 1995; see Mereghetti 2000
for a review) and the Soft Gamma-ray Repeaters (SGRs; e.g., Hurley
2000). According to the magnetar models, the quiescent X-ray emission
from SGRs and the persistent pulsed X-ray emission of the AXPs may be
powered by internal (crustal) heating arising from the decay of such
strong magnetic fields, thought to be generated by dynamo processes in
proto-neutron stars (Usov 1992; Thompson \& Duncan 1993).
Alternatively, strong magnetic fields direct the internal heat of the
young \nss\ preferentially to the magnetic poles generating an
anisotropic surface temperature distribution and may therefore give
rise to the pulsed emission of SGRs (Usov 1997) and AXPs (Heyl \&
Hernquist 1997). In addition, magnetospheric processes may contribute
significantly to the total emission from magnetars (e.g., Thompson \&
Duncan 1996) and have often been suggested as the explanation for the
hard spectral tails observed in these sources.

Recent observations have provided support for the applicability of the
magnetar models to AXPs and SGRs. The isolated nature of these
sources, suggested by the lack of observable companions (Mereghetti,
Israel, \& Stella 1998; Hulleman, van Kerkwijk, \& Kulkarni 2001), as
well as their apparent youth, indicated by the potential
supernova-remnant associations (Kaspi 2001) and their proximity to the
Galactic plane, are naturally accounted for by magnetar models. Some
of the timing properties, such as the large period derivatives and
noisy spin-down (Woods et al.\ 1999; Thompson et al.\ 2000; Kaspi,
Lackey, \& Chakrabarty 2000) may also be related to the presence of an
ultrastrong magnetic field.

In the absence of calculations addressing the angle- and
energy-dependent emission from the surface of ultramagnetized \nss,
however, it has not been possible to date to perform similar detailed
comparisons of models to observable properties such as the pulse
profiles and spectra. Analysis of folded lightcurves and spectral
modeling so far has had to rely on simple estimates and
parameterizations. Significant qualitative changes in the transport of
radiation in the presence of strong magnetic fields prohibit an
extrapolation of results from model atmospheres obtained for
nonmagnetic or weakly-magnetic \nss. Furthermore, the magnetic
opacities have a very strong dependence on photon energy and on the
angle with respect to the magnetic field direction; thus, angle- and
energy-integrated approaches to this problem are insufficient (see also
\S 3). Motivated by this, we have carried out radiative transfer 
calculations in magnetar atmospheres, for magnetic fields in the range
$10^{13} \leq B \leq 10^{15}$ G, which is a previously unexplored
regime.

Significant progress has been made recently by several authors in
modeling \ns\ atmospheres with typical ($\sim 10^{12}$~G) or
negligible magnetic fields. These calculations have been carried out
in the context of weakly-magnetic \nss\ with H atmospheres (Shibanov
et al.\ 1992; Pavlov et al.\ 1994; Zavlin, Pavlov, \& Shibanov 1996),
as well as for magnetized heavy-element atmospheres (Rajagopal \&
Romani 1996; Rajagopal, Romani, \& Miller 1997) with varying degrees
of approximation. The coherent scattering of radiation has so far been
included only in the case of weakly magnetized atmospheres with
isotropic opacities (Zavlin et al.\ 1996). In the case of magnetized
atmospheres with anisotropic opacities, angle-averaged solutions have
been obtained using the moment equations (Rajagopal \& Romani 1996,
Shibanov et al.\ 1992, Pavlov et al.\ 1994) or scattering has been
neglected altogether (Rajagopal et al.\ 1997).

The current work represents, to our knowledge, the first solution of
the strongly magnetized radiative equilibrium atmosphere problem with
angle-dependent scattering and the inclusion of vacuum polarization
effects. As the increasing magnetic field strength renders the angle
and energy dependence of the photon-electron interaction cross
sections more severe, as well as qualitatively altering the radiation
transport properties because of vacuum polarization effects, this new
regime requires a new set of tools for the radiative transfer problem.
The description of these methods as well as the results from the model
atmospheres is the subject of this paper.

The important effects of the resonance due to the polarization of the
magnetic vacuum on radiation spectra has previously been addressed by
Bulik \& Miller (1997). They show that broad-band absorption features
appear due to the enhanced interaction cross sections when high energy
photons propagate through a strongly magnetized plasma. Zane et al.\
(2000) also present a solution of the ultramagnetized problem with
scattering but suffer from the use of ordinary lambda iteration method
for solution of the scattering which is known to have convergence
problems (see, e.g., Mihalas 1978).  After the submission of the
present paper, two independent treatments of ultramagnetized
atmospheres have appeared. Ho \& Lai (2001) and Zane et al.\ (2001)
include the proton cyclotron line but neglect the vacuum polarization
resonance. It should also be noted that Ho \& Lai (2001) do not retain
the full angle dependence, while Zane et al.\ (2001) employ a local
temperature correction scheme for the radiative equilibrium
atmosphere, which does not yield radiative equilibrium solutions.

The thermodynamic properties of magnetized crusts and envelopes have
been addressed previously (e.g., Heyl \& Hernquist 1998; Potekhin \&
Yakovlev 2001).  It has been shown that heat transport is affected not
only by the surface magnetic-field strengths, but by the chemical
composition of the neutron star atmospheres as well. In particular,
large fluxes through the \ns atmosphere can be obtained only in the
presence of a light-element (H and He) atmosphere, because the heat
conductivity is inversely proportional to the atomic charge in strong
magnetic fields. Furthermore, in the case of observations of soft
X-ray emission from \nss\ and thermal optical emission from young
radio pulsars, spectral fits have required the presence of a H layer
on the \ns\ (Pavlov et al.\ 2001). It is for these reasons that, in
the current work, we restrict our attention to the transport of
radiation through a H plasma.

In \S 2 we outline the physical processes in a strongly magnetized
plasma and the equations that describe a magnetized atmosphere in
radiative equilibrium. In \S 3 we present the method of solution, in
\S 4 we discuss the energy and angle dependence of the radiation 
emerging from the \ns\ surface for a range of plasma parameters and,
in \S 5, we discuss the various implications of our results.

\section{The Problem}

In this paper, we consider the transport of radiation through a
strongly magnetized ($B \gtrsim 10^{13}$~G) atmosphere of a \ns\ and
construct radiative equilibrium models to calculate the spectrum of
the emerging radiation. We assume that the atmosphere is a fully
ionized H plasma with an ideal-gas equation of state in local
thermodynamic equilibrium (LTE). Owing to the fact that the thickness
of the atmosphere ($\approx 10$~cm) is much smaller than the radius of
the \ns, we treat the atmosphere in plane-parallel geometry. In
addition, we assume that the atmosphere has no lateral density
variations and that the magnetic field is normal to the \ns\
surface. This reduces the radiative transfer problem to one spatial
dimension which may be specified, e.g., by the column density in the
atmosphere measured from the surface or the equivalent Thomson
scattering optical depth $\tes$.

The problem, then, consists of solving the equation of radiative
transfer with scattering in a magnetized plasma, subject to the
constraint of radiative equilibrium, under the assumption of LTE,
together with the equation of hydrostatic equilibrium. Below, we first
present the processes occuring during the interaction of photons with
a magnetized plasma, as well as the approximate ionization equilibrium
of the plasma in the presence of a strong magnetic field. We then
state the equations for the model-atmosphere problem for the specific
case of an ultramagnetized \ns. Note that, throughout this paper, we
express the photon energy $E$ and the plasma temperature $T$ in keV,
and report all quantities as measured by a static observer on the
\ns\ surface.

\subsection{Properties of Strongly Magnetized Plasmas}

In this section, we briefly summarize some of the properties of
electrons and atoms in strong magnetic fields, $ B \gg m_e^2
e^3c/\hbar^3 = 2.35 \times 10^9$~G, where $m_e$ and $e$ are the
electron mass and charge, c is the speed of light, and $\hbar$ is
Planck's constant (for a thorough treatment of this subject, see
M\'esz\'aros 1992, a recent review by Lai 2001, and references
therein). We concentrate, in particular, on field strengths greater
than $B_{\rm{cr}} = 4.3 \times 10^{13}$~G, at which the cyclotron
energy of the electron $E_{\rm b} = e B / m_e c $ equals its rest
mass, 511 keV. At these magnetic field strengths and at temperatures 
$T\lesssim 1$~keV, $E_{\rm b} \gg E_{\rm F}, E_{\rm th}$ where $E_{\rm
F}$ and $E_{\rm th}$ are the Fermi energy and the thermal energy of
the electrons, respectively; hence, the electrons are confined to the
ground Landau level (in the so-called adiabatic approximation).  This
greatly limits the motion of the electrons in the transverse direction
and the electron is nearly confined to motion in one dimension along
the magnetic field. The photon-electron cross section can thus be
greatly reduced depending on the direction of polarization of the
photon with respect to the magnetic field. 

Neutral atoms can also play an important role in plasmas at high
magnetic field strengths. The binding energies of atoms are highly
enhanced so that, for $B \gtrsim 10^{12}$ G, the H atom ground-state
ionization potential is $\sim 300-500$~eV (Potekhin 1998). Therefore,
the ionization equilibrium of a plasma can differ significantly from
the nonmagnetic case and the neutral fraction even at a temperature of
1~keV can be large. A competing effect in \ns\ atmospheres is that of
pressure ionization, which dominates the ionization equilibrium at
high densities, $\rho \gtrsim 10^2 {\rm g~cm^{-3}}.$ The most complete
treatment to date of ionization equilibrium in strongly magnetized
plasmas is by Potekhin, Chabrier, \& Shibanov (1999), who used the
numerically calculated energy levels of a H atom to compute the
thermodynamic properties of a non-ideal gas and the neutral fractions,
for a range of temperatures and densities, and for magnetic field
strengths of $\lesssim 10^{13}$ G.

The present calculations span a range that is higher than the
calculations of Potekhin et al.\ (1999) in all three quantities, i.e.,
temperature, density, and magnetic field strength. Specifically, most
of the photons which carry the flux in the model atmospheres originate
at a depth with $T \approx 1$~keV and $\rho \gtrsim 10^3 {\rm
g~cm^{-3}}.$ Extrapolating the binding energies for neutral H at the
highest magnetic field strengths calculated by Potekhin (1998),
assuming that only the centered (lowest quantum-number) states are
populated due to the excluded-volume effects at these high densities,
and using the magnetic Saha equation (e.g., Lai 2001), we estimate
that the neutral fraction is at most $f_H < 10^{-5}$. Therefore, in
the present work, we make the assumption of complete ionization in the
plasma and defer a more complete treatment to a later paper. Note that
we also employ in our calculations an ideal gas equation of state
which provides a good approximation to the state of the plasma at the
temperatures and densities where $E=$~few~keV photons originate in the
atmosphere.

\subsection{Transport of Photons in Strongly Magnetized Plasmas and 
Interaction Cross Sections}

At magnetic field strengths in the range $10^{13}~{\rm G} \leq B 
\leq 10^{15}$~G which we consider here, the electron cyclotron energy
$E_b$ is much greater than the plasma temperature $kT \lesssim 1$~keV
so that the thermal motions of the electrons can be neglected (the
cold plasma approximation). At the same time, for these magnetic field
strengths, the photon energies in the range of interest ($0.1 \leq E
\leq 10$~keV) are $\ll E_b$ and therefore lie sufficiently far from
cyclotron resonances. Under these conditions, the transport of photons
through a plasma can be described by two orthogonal normal modes which
correspond to different polarization states of the photons (e.g.,
Gnedin \& Pavlov 1974; Pavlov \& Shibanov 1979). Here we adopt $i=1$
for the extraordinary mode and $i=2$ for the ordinary mode and denote
by $\ii$ the monochromatic intensity of radiation emerging from the
plasma in each mode at photon energy $E$. Note that in a pure plasma,
the normal modes correspond to right- and left-hand circularly
polarized photons for propagation angles $\theta \approx 0$, while for
$\theta \approx \pi/2$, they are linearly polarized orthogonal to and
along the magnetic field. The normal modes for all other directions of
propagation are elliptically polarized. (See \S 2.3 for the effects of
vacuum polarization on the normal modes.)

In our calculations, we take into account scattering as well as
emission and absorption due to magnetic bremsstrahlung processes.
Because the conditions $E \ll m_e c^2$ and $kT \ll m_e c^2$ are
satisfied, we neglect the change in photon energy in the scattering
process (see Gonthier et al.\ 2000 for these cross sections). The
scattering coefficient $\chi_{ij}(\mu,\mu^\prime)$ gives the
probability of scattering for a photon with direction of propagation
$\mu = \cos \theta$ with respect to the magnetic field and
polarization mode $i$ to a direction $\mu^\prime$ and polarization
mode $j$. We follow Kaminker, Pavlov, \& Shibanov (1982;
eqs.~[40]--[53]) to calculate the scattering coefficients. Note that,
as in Kaminker et al.\ (1982), the scattering and absorption
coefficients throughout this paper are given in units of $N_e
\sig$, where $N_e$ is the electron density and $\sig$ is the 
angle-averaged Thomson cross section. The total scattering coefficient
is denoted by $\chisi$ and is given by $\chisi = \sum_j \int
\chi_{ij}(\mu,\mu^\prime) {\rm d}\mu^\prime$. The absorption
coefficient $\chiai$ is related to the scattering by the generalized
Coulomb logarithms and can be found in Kaminker, Pavlov, \& Shibanov
(1983; eqs.~[A1]--[A9]).

\begin{figure}[t]
 \centerline{ \psfig{file=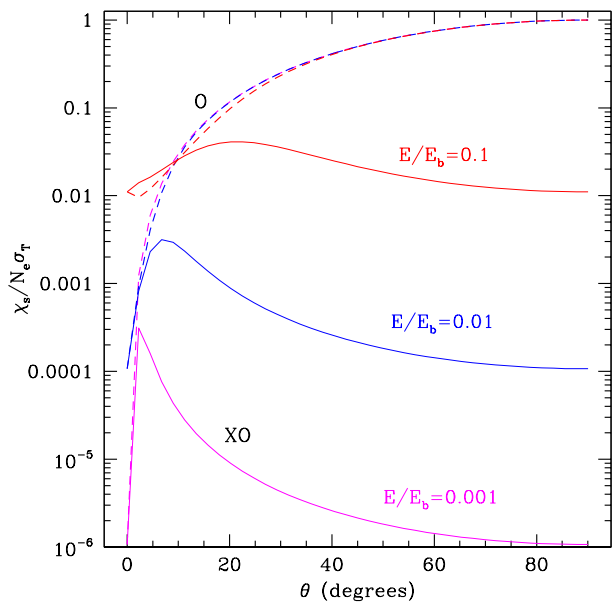,angle=0,width=8.5truecm} }
\figcaption[]{The scattering coefficient $\chisi$, 
in units of $N_e \sig$, for the extraordinary (solid lines) and the
ordinary (dashed lines) mode as a function of $\theta$ for $E/E_b =
0.1, 0.01$, and $0.001$. \label{Fig:opac}}
\end{figure}

The magnetic scattering and absorption coefficients have strong
dependences on the direction of propagation and photon energy, and the
two modes show increasingly different behavior with increasing
magnetic field strength. To facilitate the discussion in $\S 4$ and
make our results more intuitive, we plot in Figure~\ref{Fig:opac} the
scattering coefficients $\chisi$ as a function of $\theta$ for $E/E_b
= 0.1$, $0.01$, and $0.001$. The solid lines show the extraordinary
and the dashed lines show the ordinary-mode opacities. In $\S 4.3$, we
will show that the beaming of emerging radiation for different values
of $E/E_b$ is closely related to the behavior of the opacities shown
in Figure~\ref{Fig:opac}.

\subsection{Vacuum Polarization}

In strong magnetic fields, in addition to the response of the plasma
electrons, the vacuum polarization effects due to the virtual
electron-positron pairs also become significant (Adler 1971; Gnedin et
al.\ 1978; M\'esz\'aros \& Ventura 1978, 1979) and contribute to the
dielectric tensor and the magnetic permeability tensor. Thus, the
vacuum polarization alters the polarizations of the normal modes and
affects the interaction cross sections in the plasma (see Pavlov \&
Shibanov 1979; M\'esz\'aros \& Ventura 1979; M\'esz\'aros 1992).  The
effect of the magnetic vacuum can be quantified by the parameter 
\be 
W = \left( \frac{3 \times 10^{28} {\rm cm}^{-3}}{N_e}\right) \left(
\frac{B}{B_{cr}}\right)^4.  
\ee 
which enters the ellipticity parameter $q$ of the normal modes through
\be q=\frac{\sin^2\theta}{2\cos\theta} \sqrt u
\left(1-W\frac{u-1}{u^2}\right), \ee where $u=E_b^2/E^2$.  The
resulting normal modes of the magnetic vacuum, in the absence of
plasma electrons, are linearly polarized at $\theta=0$ and
$\theta=\pi/2$, and elliptically polarized at all other angles of
propagation. Note that the ellipticity parameter $q$ has a slightly
modified form for $B>B_{cr}$ but this does not affect qualitatively
the emerging spectrum (\"Ozel \& Narayan 2001).

\begin{figure}[t]
\centerline{ \psfig{file=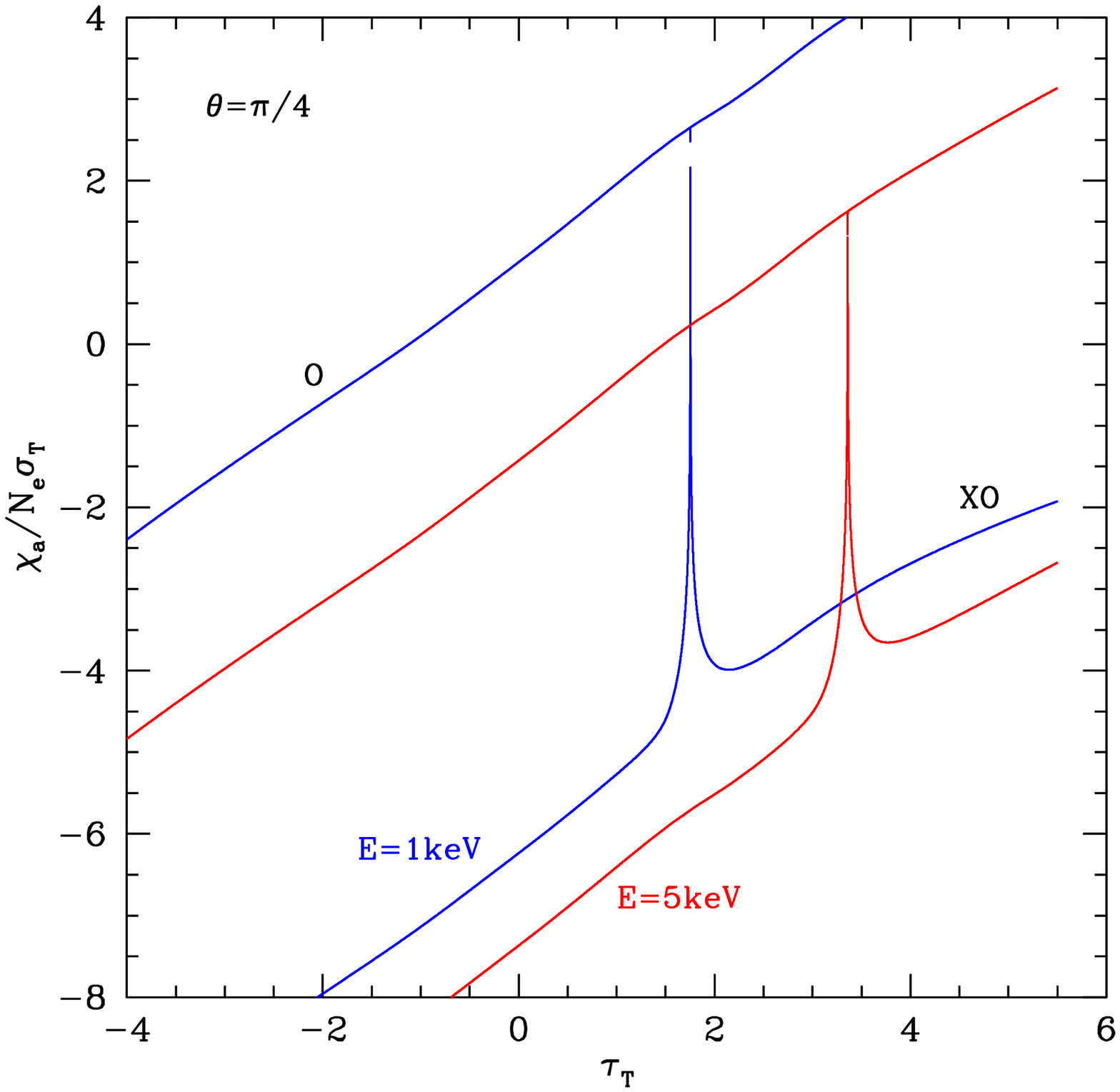,width=8.0truecm} } 
\centerline{\psfig{file=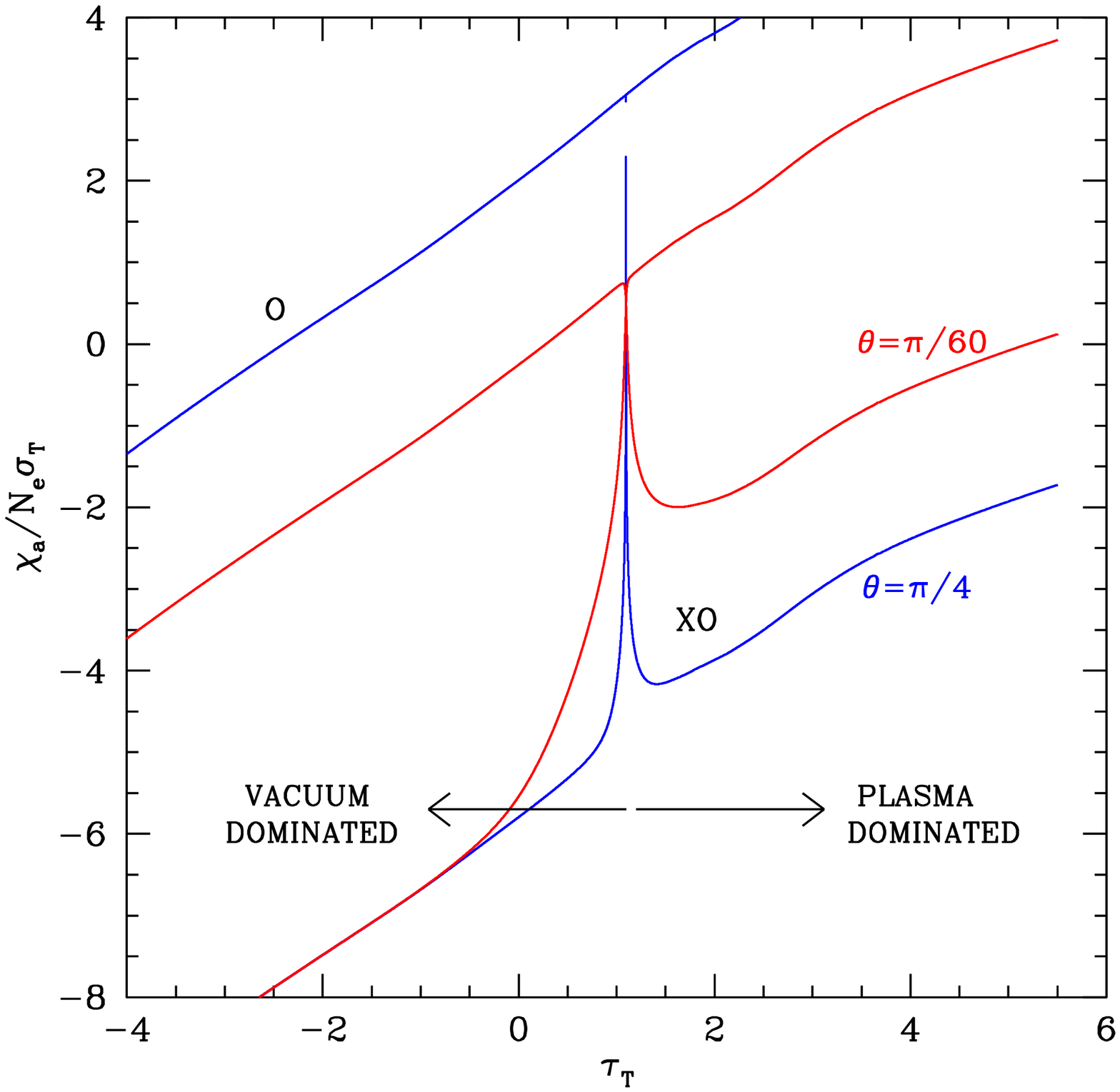,width=8.0truecm} } 
\figcaption[]{The absorption coefficient $\chiai$, in units of 
$N_e \sig$, of the extraordinary and ordinary modes as a function of 
density in the atmosphere in the presence of vacuum polarization at 
two photon energies (top panel) and for two directions of propagation 
at $E=0.5$~keV (bottom panel). \label{Fig:vac}}
\end{figure}

When both plasma electrons and virtual pairs are present, the normal
modes of propagation at very high plasma densities (plasma-dominated
region) are different than those at very low plasma densities
(vacuum-dominated region) and go through a transition at a critical
density where plasma electron and virtual pair contributions
identically cancel out (see Fig.~\ref{Fig:vac}). At this critical
density, a resonance appears at the point where $q$ changes sign and
the linear polarizations of the normal modes vanish, leaving the modes
purely circularly polarized. (Strictly at the critical frequencies,
the orthogonality of the normal modes also vanishes, rendering the
normal mode treatment invalid; see M\'esz\'aros \& Ventura 1979 for a
discussion). For $W > 4$, the $q=0$ condition can be written as \be
E^2 = 0.5 E_b^2 \left[1 \pm (1-4/W)^{1/2}\right], \ee which gives the
critical photon energies for each density where this resonance
appears. Alternatively, the same equation can be used to determine the
critical density at which the resonance occurs for each photon energy
$E \leq E_b$.

The most important effect of the vacuum resonance is that, for each
photon energy, the opacities of the two modes have a close approach or
become equal at the critical density depending on the value of $\mu$,
with a strong enhancement of both the scattering and absorption
coefficients of the extraordinary mode, including the off-diagonal,
mode-changing terms of the scattering matrix. Figure~\ref{Fig:vac}
shows the opacities with vacuum resonance as a function of
$\tau_T=\int N_e \sig dz$ in the atmosphere (see also \S 3.1).  The
resonance occurs over a narrow range of densities, which leads to
broadband absorption-like features in the spectrum (see \S 4.3 and
also Bulik \& Miller 1996). Finally, the eigenmodes of the pure vacuum
are similar to those of the electron plasma for most angles $\theta$,
but differ significantly at a very small range of angles near
$\theta=0$. Therefore, even though an extraordinary-mode photon can
convert into an ordinary-mode photon across the resonance through
scattering, the definitions of the ``extraordinary'' and the
``ordinary'' modes essentially remain the same above and below the
resonant density (see also Fig.~\ref{Fig:vac} and M\'esz\'aros 1992).

The vacuum resonance has a finite narrow width in photon energy and
electron density. For a given photon energy and direction of
propagation, the width of the resonance corresponds to a change in
$q$, of \be \Delta q \simeq \frac{\sin^2\theta}{2\cos\theta}
\frac{E_b}{E} \frac{\Delta N}{N}, \ee and occurs over a fractional
change in electron density of $\Delta N/N \lesssim 1\%$. We include in
our calculations the effects of vacuum polarization through the normal
mode ellipticity parameter, as given in Kaminker et al.\
(1982). However, due to the numerical resolution which prevents us
from sampling the change in electron density to the required accuracy
even with 400 depth points, we force the resonance to saturate by
choosing for $\vert q \vert$ a minimum allowed value across the
resonance of \be \vert q \vert \ge \xi
\frac{\sin^2\theta}{2\cos\theta} \frac{E_b}{E}\,.  \ee We have
experimented with the factor $\xi$ and found that for our resolution,
$\xi \sim 1\%$ results in smooth spectra with negligible change in the
total and scattering optical depth. This is equivalent to assuming
that at a given $\tes$, the density fluctuates to within $\sim 1\%$.

Our results show that specifically, for $B > 10^{14}$~G and at
densities $N_e \lesssim 10^{27} {\rm cm}^{-3}$, $W \gg 1$ and
therefore the temperature profiles of the atmospheres as well as the
emerging spectra at energies above 1~keV are significantly modified by
vacuum polarization (see \S 4.1 and 4.3). Already based on
Figure~\ref{Fig:vac}, one can see that this resonance is expected to
attenuate the emerging flux and lead to a heating of the atmosphere at
the $1 \lesssim \tau_{\rm T} \lesssim 10^4$, which corresponds to the
density range that includes the critical density for most flux
carrying ($0.5-5$~keV) photons.

\subsection{Model Atmospheres}

We choose as the dimensionless depth variable the Thomson optical
depth, $d\tes = N_e \sig dz$, and define a logarithmic grid of
typically 100-400 equally spaced points between $10^{-5}$ and
$10^5$. In terms of this stratification parameter, the equation of
transfer in the two modes can be written as
\begin{eqnarray} 
y_G \mu \frac{d\ii}{d\tes} &=& \chiai \left[\ii-\frac{\bnu}{2}\right] 
+ \chisi \ii\nonumber\\
& & -
\sum_j\int \ascatt \ij(\mu^\prime) d\mup \;,
\end{eqnarray}
where $\bnu$ is the Planck function, $\chiai$ and $\chisi$ denote the
total absorption and scattering coefficients in Thomson units ($N_e
\sig$), respectively, and $\scatt$ is the angle- and
polarization-dependent scattering coefficient. The parameter $y_{\rm
G}$ takes into account the effects of general relativity and is
defined as
\be
y_{\rm G}\equiv \sqrt{1-\frac{2GM_{\rm NS}}{R_{\rm NS} c^2}}\;,
\ee
where $G$ is the gravitational constant, $c$ is the speed of light,
and $M_{\rm NS}$ and $R_{\rm NS}$ are the neutron-star mass and
radius, respectively. In the following calculations, we set $M_{\rm
NS}= 1.4 M_\odot$ and $R_{\rm NS}=10^6$~cm.  Since there are no
sources or sinks of heat, the atmospheres are in radiative
equilibrium, specified by
\be 
H_{\rm eff} = \frac{1}{2} \sum_i\int\int \ii(\mu)\mu d\mu dE
\equiv \frac{1}{4\pi} \sigma T_{\rm eff}^4\;,
\ee 
where $4\pi H_{\rm eff}$ is the (constant) radiation flux and
$T_{\rm eff}$ is the effective temperature.

The momentum balance in the atmosphere is specified by the 
hydrostatic equilibrium condition 
\be
\label{eq:hy}
\frac{dP}{d\tes} = \frac{g \rho}{y_G^2 N_e \sig}\;,
\ee
where $P$ is the pressure, $\rho$ is the density, and $g=GM_{\rm
NS}/R_{\rm NS}^2$. This system of equations describe the state of the
plasma and can be closed assuming an equation of state, which to first
approximation can be taken to be that of an ideal gas (see \S2.1),
\be
\label{eq:eos}
P = 2 N_e k T\;,
\ee
where the factor of 2 is valid for completely ionized H gas. 
Equations~(\ref{eq:hy}) and (\ref{eq:eos}) can be easily integrated
from the surface inwards, to give
\be
\label{eq:dens}
N_{\rm e} = \frac{gm_{\rm p} \tes}{2 y_G^2 \sig k T}\;,
\ee
where $m_{\rm p}$ is the proton mass. The temperature and density
profile of the atmosphere is obtained by solving these equations
coupled with the radiation field. The iterative method of solution for
this set of equations is described in the next section.

\section{The Method}

We compute the radiative equilibrium model atmospheres using an
iteration procedure that involves the solution of the equation of
radiative transfer through a Feautrier method and makes use of the 
linearized moment equations to determine the temperature profile. 
 
\subsection{Radiative Transfer}

We follow a modified Feautrier method (Mihalas 1978) for the solution
of the radiative transfer problem. This method allows for a complete,
non-iterative solution (of the transfer equation), including the
effects of angle-dependent scattering and absorption cross
sections. Because we need only to consider conservative scattering, we
solve the transfer equation at each photon energy separately.

We begin by writing the equation of radiative transfer for the 
two modes as 
\be
\label{eq:us}
y_{\rm G}\mu\frac{\dii}{\dtaui} = \ii-\si\;,
\ee
where 
\be
\label{eq:tau}
d\tau^i \equiv (\chiai+\chisi) d\tau_{\rm T}
\ee
is the total optical depth at each mode and we have defined the
energy-dependent source function as
\be
\label{eq:sf}
\si \equiv \frac{\chiai}{\chiai+\chisi} \frac{B}{2} + \frac{1}
{\chiai+\chisi} \sum_j\int \scatt \ij(\mu^\prime) d\mup \;.
\ee
In the following equations we drop the energy subscript and explicitly
specify any energy-integrated quantities.

We define the Feautrier variables
\be 
\label{eq:u}
\ui \equiv \frac{\iip + \iim}{2}
\ee
and
\be 
\label{eq:v}
\vi \equiv \frac{\iip - \iim}{2}.
\ee
We then write the transfer equation~[\ref{eq:us}] as two first-order
equations
\be
\label{eq:ft1}
y_{\rm G}\mu \frac{\partial \ui}{\dtaui} = \vi
\ee
and 
\be
\label{eq:ft2}
y_{\rm G}\mu \frac{\partial \vi}{\dtaui} = \ui-\si\;,
\ee
which can be combined to give one second-order equation for the 
variable $u$
\be
\label{eq:ft}
y_{\rm G}^2\mu^2\frac{\partial^2 \ui}{{\dtaui}^2} = \ui-\si\;.
\ee

Note here that, unlike the usual Feautrier equations, in which the
source function involves an integral over only photon momentum, the
source function in equation~[\ref{eq:ft}] involves also a summation
over the photon polarization mode, introducing a coupling between the
two equations. However, the off-diagonal (mode-changing) elements of
the scattering coefficients, $\scatt$ ($i\ne j$), are in general much
smaller than the diagonal elements away from the cyclotron resonance,
and can be treated as a perturbation.  Making use of this fact, we
solve iteratively the Feautrier equation for each mode separately,
assuming that the coupling terms are known from the previous
iteration, until the local specific intensity converges to one part in
$\sim 10^{9}$.

The transfer problem is highly coupled in the three variables
$\tau_{\rm T}, E$, and $\mu$ so that choosing a discretization grid is
one of the most delicate parts of the calculation. We discretize the
second-order equation~(\ref{eq:ft}) on the predefined grid over the
variable $\tes$ (and not over $\taui$, which depends on $\mu$ and
$E$), according to Mihalas (1978; eq.~[6.26]-[6.30]). We use
\be
\Delta\tau_{i,d\pm1/2}\equiv \frac{1}{2}(\chi^{\rm tot}_{i,d}
 +\chi^{\rm tot}_{i,d\pm 1})
 \vert \tau_{\rm T,d}-\tau_{\rm T,d\pm 1}\vert
\ee
where $\chi^{\rm tot}_i\equiv\chiai+\chisi$. The chosen range for
$\tes$ (see \S2.3) corresponds to a range in the optical depth in the
extraordinary mode ($\tau_1$) of $0.0001-10$ (depending on the photon
energy and direction of propagation) and in the ordinary mode
($\tau_2$) of $0.001-10^4$. We chose this range so that all photon
energies of interest turn optically thick in the deep interior and
decouple from the atmosphere at small optical depths.

Because of the very strong dependence of magnetic opacities on angle
near $\mu \simeq 1$, we choose a logarithmic grid (typically of 30--40
points) in the quantity $1-\mu$, with $\log(1-\mu) = -5\rightarrow 0$.
We also choose a logarithmic grid (typically of 40 points) in photon
energy, with $\log E = -3\rightarrow 1.5$.  This range spans the
photon energies that carry most of the radiative flux for the
effective temperatures ($\sim 0.1-0.5$~keV) of our atmosphere.  Note
that the highest number of points in depth, photon energy, and angle
are required in those calculations where the vacuum polarization
effects are significant.

Finally, we follow the procedure described in Mihalas (1978) to
evaluate and invert the Feautrier matrices. 

\subsection{Temperature Correction Scheme}

We start with an initial guess for the temperature profile
\be
\label{eq:tinit}
T = \left[\frac{3}{4}
\left(\tau_{\rm R}+\frac{2}{3}\right)\right]^{1/4} T_{\rm eff}\,
\ee
where $\tau_{\rm  R}$ is the  optical  depth  that corresponds to  the
Rosseland  mean opacity $\chi_{\rm R}$, defined as (Pavlov et al.\
1992)
\be
\label{eq:taur}
\frac{1}{\chi_{\rm R}} = \frac{3 \pi}{8 \sigma T^3}
\int_{-1}^1 d\mu \int dE \left[
   \frac{1}{\chi^{\rm tot}_1}+\frac{1}{\chi^{\rm tot}_2}\right]
\frac{\partial B_E(T)}{\partial T}
\ee
for two polarization modes and calculate the corresponding density
$N_e(\tes)$ using equation~(\ref{eq:dens}).  We then proceed by
calculating the magnetic opacities with this guess temperature profile
on our predetermined grid and solve the radiative transfer problem.

In order to determine the temperature profile of the model \ns\
atmosphere in radiative equilibrium, we use an iterative temperature
correction scheme similar to the Uns\"old-Lucy algorithm, which
utilizes the energy-integrated moment equations in order to obtain a
linear perturbation equation for $\Delta T(\tes)$ (see Mihalas 1978).
However, despite the similarity of the approach, the scheme we describe
here differs from the Uns\"old-Lucy scheme in two important
respects:

(i) We do not make any assumptions about the angular distribution of
the radiation field, which is the case when the Eddington
approximation is used. Instead, we calculate at each iteration the
Eddington factor $f = K / J$ as a function of $\tes$, as well as the
ratio $K/H$ at the boundary $\tes = 0$, where $J, H$, and $K$ denote
the zeroth, first, and second moments of the specific intensity,
respectively. This takes into account the deviations from anisotropy
in the radiation field which are likely to be present in strongly
magnetized plasmas.

(ii) We include first-order corrections in {\em all\/}
temperature-dependent quantities due to a temperature change $\dt$,
except for the radiation intensity itself. This is in contrast to the
assumption in the Uns\"old-Lucy procedure, that only the local
blackbody source function changes with temperature.

With these modifications, the correction scheme proceeds as follows.
We start with the energy-integrated zeroth and first moment equations
\be
\label{eq:mom1}
y_G \frac{\partial H}{\dtes} = \chi_J J - \chi_P B
\ee
and
\be
\label{eq:mom2}
y_G \frac{\partial K}{\dtes} = \chi_F H \;,
\ee 
where $\chi_J, \chi_P$, and $\chi_F$ are similar to (Mihalas 1978) the
absorption mean, Planck mean, and flux mean opacities, respectively,
appropriately defined for a magnetized atmosphere. In particular
\be
\chi_P\equiv \frac{\pi}{2}\left[\frac{\int d\mu \int dE B_E(T)
   (\chi^a_1+\chi^a_2)/2}{\sig T_{\rm eff}^4}\right]\;,
\ee
\be
\chi_J\equiv \frac{\int d\mu \int dE 
   (\chi^a_1 u^1 + \chi^a_2 u^2)}{\int d\mu \int dE (u^1+u^2)}\;,  
\ee
and
\be
\chi_F\equiv \frac{\int d\mu \int dE \mu
   (\chi^{\rm tot}_1 v^1 + \chi^{\rm tot}_2 v^2)}
{\int d\mu \int dE \mu (v^1+v^2)}\;.
\ee
In the above definitions, we implicitly assume that the moments and
the mean opacities depend on $\tes$ and omit any explicit references.

Integrating equation~(\ref{eq:mom2}) and substituting into
equation~(\ref{eq:mom1}) by making use of the Eddington factors, which
are known from the current iteration, yields
\be
\label{eq:eqn3}
\chi_P B = \chi_J \left(\frac{J}{K}\right) \left[ \left(\frac{K}{H}
\right)_0 H(0) + \frac{1}{y_G} \int \chi_F(\tes^\prime)H(\tes^\prime)
d\tes^\prime \right] - y_G \left(\frac{dH}{d\tes}\right),
\ee
where the subscript 0 corresponds to the outer boundary.  We assume
that the constant flux solution can be obtained by making a small
correction $\dt(\tes)$ everywhere in the atmosphere and allow for the
local black body and the mean opacities to vary linearly with $\dt$,
which gives $\Delta B(\tes) = (\partial B / \partial T)
\Delta T \equiv B^\prime(\tes) \dt, \Delta \cp = \cp^\prime \dt$, 
and similarly for the other mean opacities. Note that all the
derivatives are computed at the $\tes$ grid points. Now proceeding as
in the Uns\"old-Lucy scheme, we note that in the desired radiative
equilibrium solution, the term $\partial H/\dtes$ vanishes and the
radiation flux becomes $H(\tes) = H_{\rm eff} = \frac{1}{4\pi}\sig
T_{\rm eff}^4$.  Thus, subtracting the current equation from the
desired constant flux solution and solving for $\dt$ we obtain
\be 
\dt(\tes) = \frac{\cj (J/K)_\tau \left[(K/H)_0 \dh(0) + I_1 \right] +
y_G ({\rm d}H/{\rm d}\tau)_\tau}
{\chi_P^\prime B + \cp B^\prime - \cj^\prime (J/K)_\tau \left(K/H)_0
H_{\rm eff} + I_2 \right] }, 
\ee 
where  
\be  
I_1 = \frac{1}{y_G} \int_0^\tau [\cfp \dh(\tesp) + \cfpp \dtp 
H_{\rm eff}] \dtesp\,, 
\ee 
\be
I_2 = \frac{1}{y_G} \int_0^\tau [\cfp + \cfpp \dtp]
H_{\rm eff} \dtesp\,,
\ee
and $\Delta H(\tes) = H_{\rm eff} - H(\tes)$ refers to the deviation
of the current solution from the constant flux solution at any depth
in the atmosphere. To quantify this deviation, we define the error $e
= [H_{\rm eff} - H(\tes)]/H_{\rm eff}$ and quote the convergence of
the obtained solutions in terms of this quantity.
 
Due to the assumption of linear changes in various quantities in
response to a temperature correction $\dt$, care must be taken when
applying large temperature corrections which may violate this
assumption. To this end, we have introduced a damping factor to
control the actual applied change everywhere in the atmosphere. We
experimented with damping factors that depend on the magnitude of
$\dt(\tes)$ and found that reducing the $\dt$ obtained from the above
scheme typically by $0.3-0.5$, even though somewhat slowing down the
convergence rate, results in a marked increase in the stability of the
algorithm. With this method, we achieve flux convergence down to a
maximum error of $e \approx 5\%$ in $O(10)$ iterations, improving it
to everywhere better than $1 \%$ in $O(100)$ iterations.

Comparing the scheme described here with the Uns\"old-Lucy algorithm,
we note that the correction in temperature at any point in the
atmosphere, $\dt(\tes)$, depends not only on the errors in the flux
everywhere in the atmosphere outward of that location (as in the
Uns\"old-Lucy scheme) but also on the temperature corrections applied
at all other points. This allows for stronger coupling between all
points in the atmosphere and leads to a more uniform convergence. At
the same time, convergence is much more stable, and artificial
temperature inversions that commonly arise when applying the
Uns\"old-Lucy scheme to magnetized atmospheres are not present.

\begin{figure}[t]
\centerline{ \psfig{file=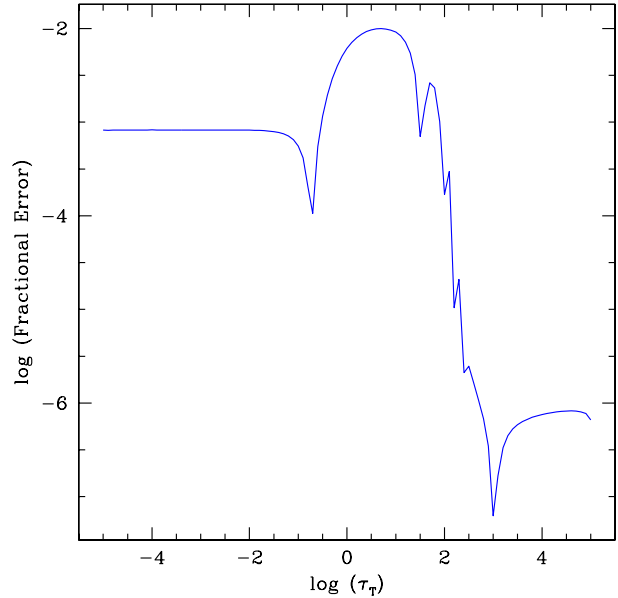,width=8.5truecm} }
\figcaption[]{The deviation from a constant flux solution throughout 
the \ns\ atmosphere. See the text for the definition of the flux
error. The parameters are $B=10^{14}$~G and $T_{\rm eff} =0.3$ keV,
chosen as a typical example. \label{Fig:err}}
\end{figure}

Finally, note that any temperature correction scheme that allows for a
change in the intensity field with temperature, such as the complete
linearization schemes (Mihalas 1978), couples the different photon
energies in the solution of the radiative transfer problem and thus
requires $O(N_E^3)$ operations in the Feautrier method, where $N_E$ is
the number of grid points in photon energy. As a result, the total
computational time required for solving the radiative transfer problem
scales as $O(M^3 N_E^3)$, with M denoting the number of points on the
$\mu = \cos(\theta)$ grid, thus limiting the computational resolution
in one or both of these variables. In such a highly angle- and photon
energy-dependent problem as the magnetized radiative transfer, we have
opted to work with the largest number of grid points in a feasible
amount of computational time and therefore chose to work with the
temperature correction scheme discussed above.

We apply the method described above to construct magnetized, radiative
equilibrium atmospheres with a flux $H(\tau)$ that is converged to
within $1\%$ of $H_{\rm eff}$, everywhere in the atmosphere.
Figure~\ref{Fig:err} shows the magnitude of typical errors that result
from such a calculation. Throughout most of the atmosphere, the
fractional error is $e \ll 1\%$, reaching $10^{-8}$ in the deep
interior. We have found that the slowest convergence, and thus the
maximum error, occurs typically in the transition region, where the
photons carrying the largest flux become optically thin; we will
discuss this transition region below. We also note that the
convergence criterion of $1\%$ does not arise from an intrinsic
limitation of the algorithms, but is chosen in view of the required
computational time.

\begin{figure}[t]
\centerline{ \psfig{file=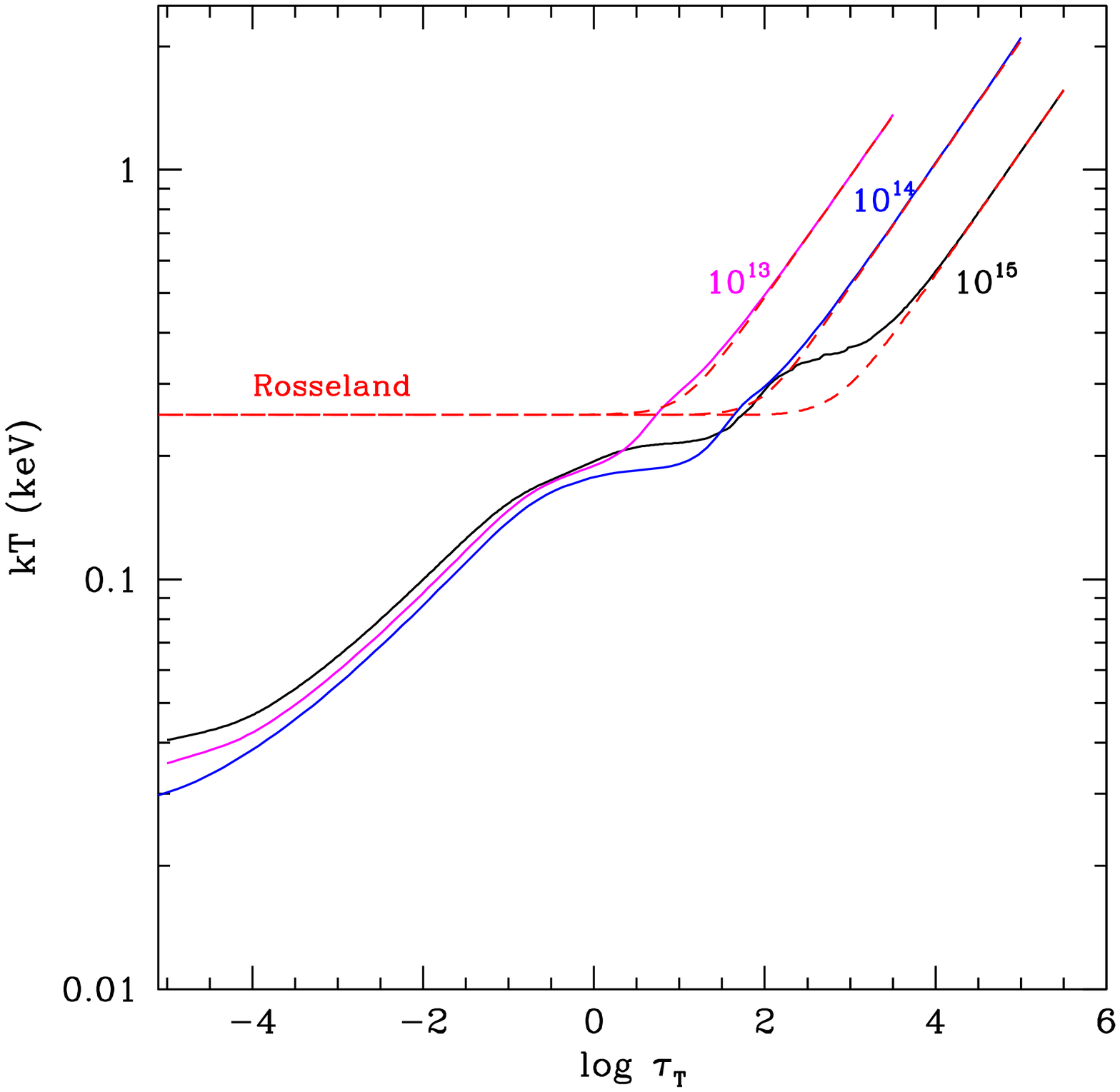,angle=0,width=8truecm} }
\centerline{ \psfig{file=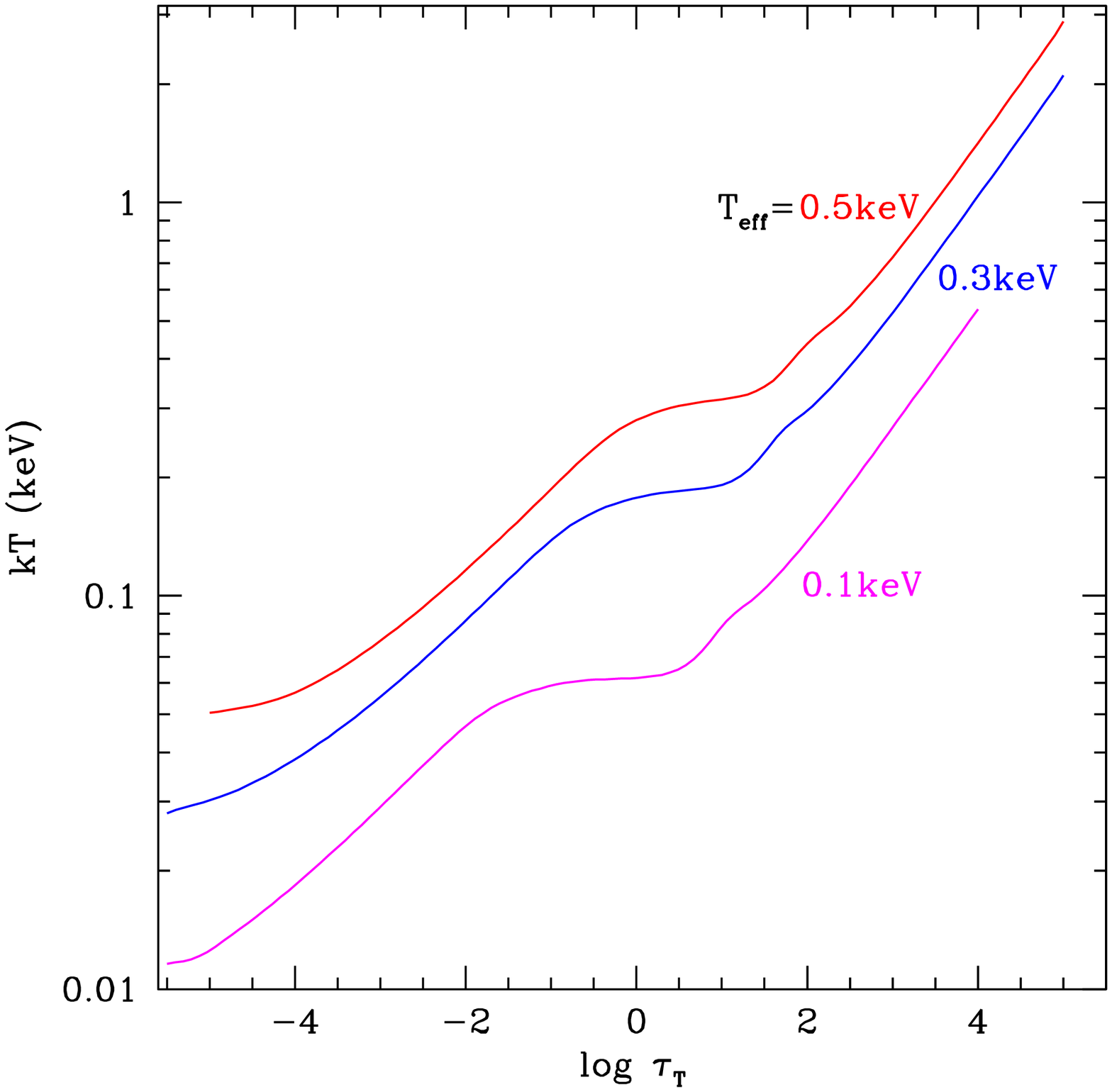,angle=0,width=8truecm} }
\figcaption[]{(Top panel) The temperature profiles
of three model atmospheres plotted against $\tes$, at $B=10^{13},
10^{14}$, and $10^{15}$ G, with $T_{\rm eff} = 0.3$ keV. The dashed
lines correspond to the temperature profiles calculated using the
Rosseland mean opacities. (Bottom panel) The variation of the
temperature profiles with $T_{\rm eff}$ for magnetic field strength
fixed at $B=10^{14}$~G. \label{Fig:temp}}
\end{figure}

\section{Results}

We have applied the method described above to construct magnetized,
radiative equilibrium atmospheres of \nss\ with a range of magnetic
field strengths ($10^{13}$~G$\le B \le 10^{15}$~G) and effective
temperatures ($0.1$~keV$<T_{\rm eff}<0.5$~keV). We show below the
resulting temperature profiles, spectra, and beaming of emerging
radiation and identify the effects of magnetic field strength,
effective temperature, scattering, and vacuum polarization on each of
these properties.

\subsection{Temperature Profiles}

Figure~\ref{Fig:temp} shows the temperature profiles for three model
atmospheres with different magnetic field strengths, at $T_{\rm
eff}=0.3$~keV (top panel). These profiles have distinct features
arising from the strong angle- and energy-dependence of the magnetic
opacities and the presence of two polarization modes and are
significantly different from the temperature profile of a grey
atmosphere (see, e.g., Mihalas 1978) and from an atmosphere calculated
using Rosseland mean opacities (dashed lines). In particular, at low
optical depths, the temperature is determined primarily by the
ordinary-mode opacities, which depend very weakly on the magnetic
field strength, for most directions of propagation.  Therefore, the
temperature profiles in this region are similar in all three model
atmospheres (but see below for the role of vacuum polarization in this
region).  This is also the reason why the Rosseland mean opacities,
which are dominated by the small opacity (extraordinary) mode, fail to
describe the temperature of the outer atmosphere correctly. At the
limit of high optical depth, the diffusion limit is reached and the
temperatures calculated using the Rosseland mean opacities provide an
accurate description of the atmospheres. At these high optical depths,
it is the extraordinary-mode opacities, which are low and depend
strongly on the field strength, that determine exclusively the
temperature profiles. As a result, with increasing magnetic-field
strength, the same plasma temperature is reached deeper in the
atmosphere.

The presence of multiple plateaus in the temperature profiles
correspond to optically thick/thin transitions of photons with
different energies, directions of propagation, and polarizations.
Specifically, the plateaus at $\tau_{\rm T}\sim 10^0-10^2$ arise from
such transitions of extraordinary-mode photons that carry most of the
flux, while the plateaus at $\tau_{\rm T}\lesssim 10^{-4}$ are the
result of such transitions for the ordinary-mode photons with low
energies or large angles of propagation.

One other significant difference between the profiles of
Figure~\ref{Fig:temp} and the temperature profile of a grey/Rosseland
atmosphere is the relation between the effective temperature and the
plasma temperature $T_0$ at the outer boundary. For a grey atmosphere,
$T_0\simeq 0.8 T_{\rm eff}$, whereas for the calculations shown in
Figure~\ref{Fig:temp}, $T_0\simeq 0.1 T_{\rm eff}$. The same
conclusion is reached by Shibanov et al.\ (1992), but the agreement is
only qualitative due to the insufficiency of the diffusion
approximation employed in that work in capturing the details of the
temperature profiles.

\begin{figure}[t]
\centerline{ \psfig{file=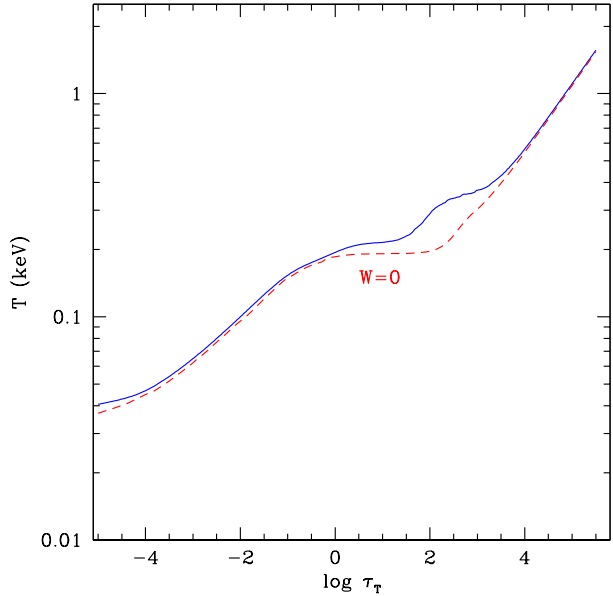,width=8.5truecm} }
\figcaption[]{The difference in the temperature profiles when the 
vacuum effects are included (solid line) and neglected (dashed line)
at $B=10^{15}$~G. \label{Fig:tvac}}
\end{figure}

The temperature profiles of model atmospheres with different effective
temperatures, as shown in Figure~\ref{Fig:temp} (bottom panel), have
the same qualitative characteristics, but show an overall
displacement. In addition, the plateaus occur further out in the
atmosphere with decreasing effective temperature, as the absorption
cross sections increase with decreasing temperature, moving the
optically thick/thin transitions to smaller column densities.

At $B \sim 10^{15}$~G, the temperature profiles are significantly
affected by the vacuum polarization resonance. To demonstrate the
vacuum resonance effects on the temperature profile, we show in
Figure~\ref{Fig:tvac} the results of two calculations at $B=10^{15}$~G
and $T_{\rm eff}=0.3$~keV, where we include (solid lines) or neglect
(dashed-lines) the vacuum polarization contribution to the magnetic
opacities. The temperature is higher for a range of optical depths $1
\lesssim \tes \lesssim 10^3$ in the case where the vacuum polarization
contribution to the opacities are included. Such an increase in
temperature is typical of radiative equilibrium atmospheres with a
resonant layer at low densities for reasons discussed below.

The net effect of the vacuum contribution to the magnetic opacities is
to increase the extraordinary-mode opacity for a narrow, photon
energy-dependent range of densities in the atmosphere (see \S 2.2 and
Fig.~2). For $B \sim 10^{15}$~G, this resonance appears at $1 \lesssim
\tes \lesssim 10^3$. Because this range lies within the photosphere
where the extraordinary mode photons of $0.5-10$~keV have decoupled
from the atmosphere, the resonance leads to an attenuation of the flux
at these photon energies, which appears as a broad-band absorption
feature in the spectrum (see \S 4.3). At the same time, the additional
absorption of radiation heats the atmosphere, thus leading to an
increase in the temperature at this range of column densities
(Fig.~\ref{Fig:tvac}). This in turn increases the emission at these
column depths and contributes to the flux at lower photon energies.
We note that it is also possible to understand these results in terms
of the radiative equilibrium condition in the atmosphere. In the
presence of resonant ``layers'' of high opacity, the overall
temperature in the atmosphere has to increase in order for the
energy-integrated flux to be constant throughout the atmosphere, for a
given effective temperature. This change in the temperature profile
results in a corresponding change in the entire emerging spectrum at
the highest magnetic field strengths, as we will show in the next
section.

\subsection{Spectra of Emerging Radiation}

\begin{figure}[t]
\centerline{ \psfig{file=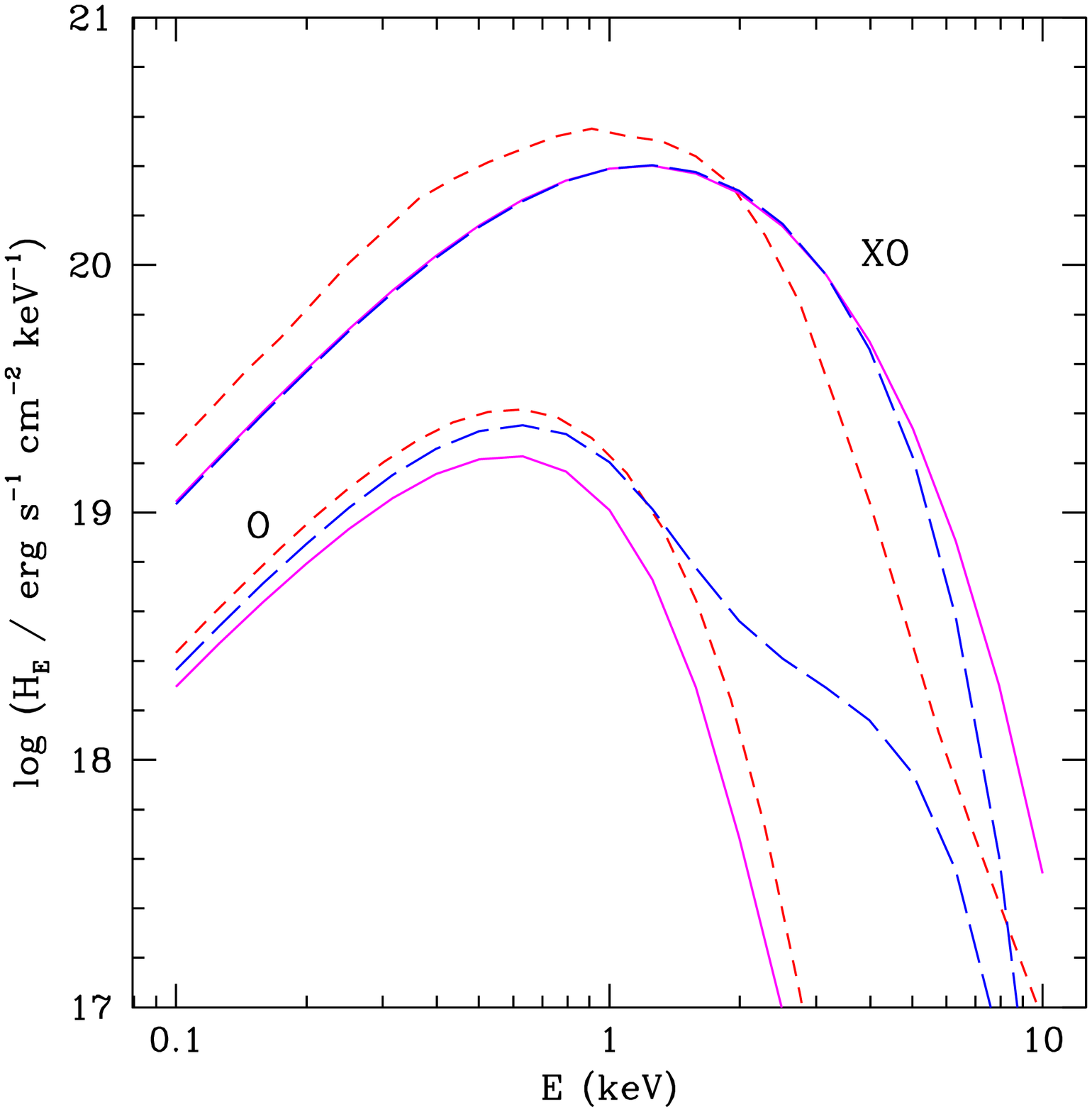,width=8.0truecm} }
\centerline{ \psfig{file=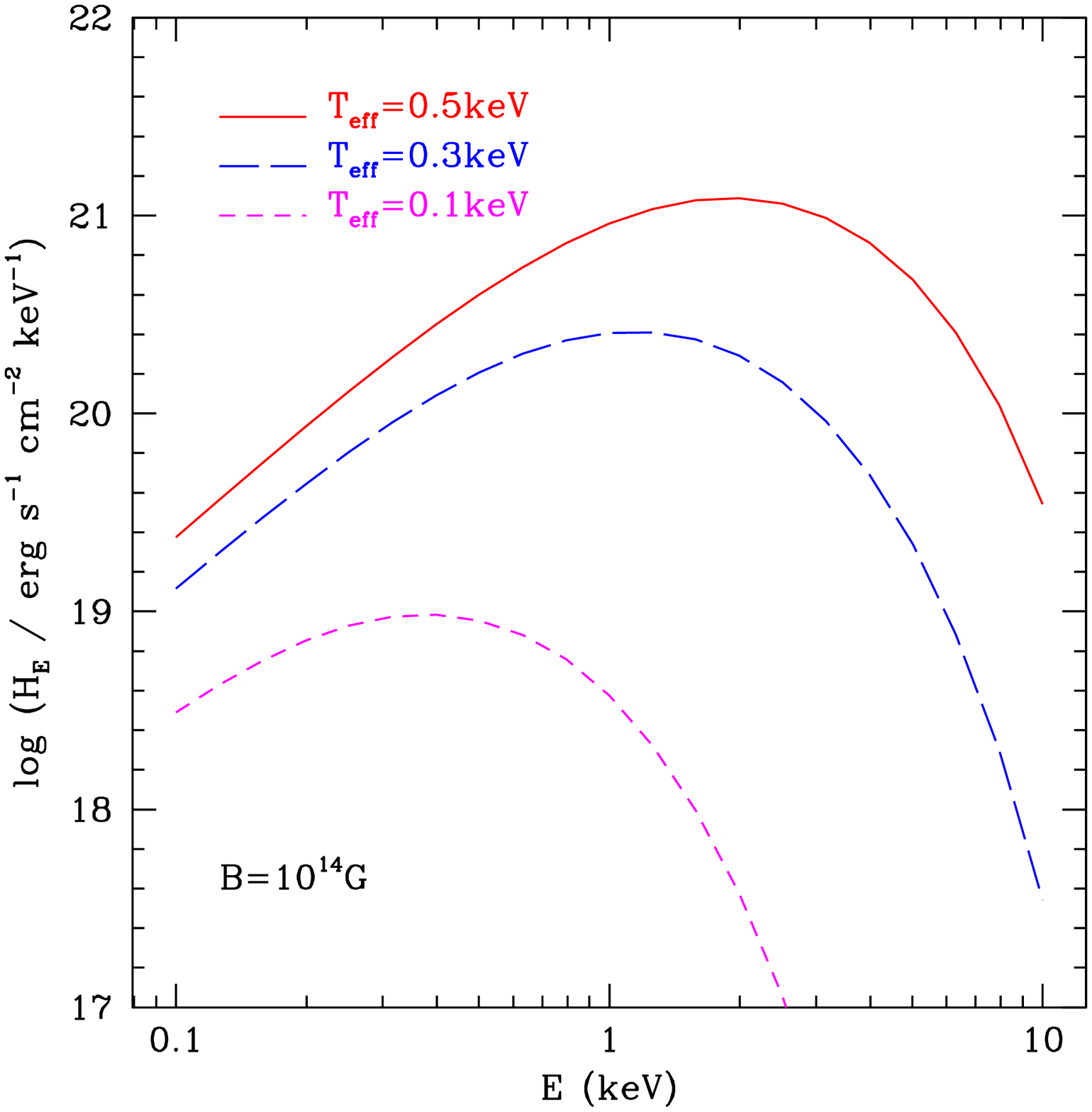,width=8.0truecm} }
\figcaption[]{(top panel) Spectra of emerging radiation at $B=10^{13}$~G 
(long-dashed line), $10^{14}$~G (solid line), and $10^{15}$~G
(short-dashed line) for $T_{\rm eff}=0.3$~keV. The flux in the two
modes are shown separately; the labels `O' and `XO' refer to the
ordinary and the extraordinary modes, respectively. (bottom panel)
Same as above, but for a fixed magnetic field strength of
$B=10^{14}$~G and for $T_{\rm eff}=0.1$~keV (short-dashed line),
$0.3$~keV (long-dashed line) and $0.5$~keV (solid line). Here, the
contributions of the two modes are added and the total flux is
plotted. The radiative flux is equal to $4 \pi H_E$. \label{Fig:spec}}
\end{figure}

In Figure~\ref{Fig:spec}, we show the spectra emerging from \ns\
atmospheres with different magnetic field strengths and effective
temperatures. In the top panel, the flux is decomposed into the
relative contributions of the two polarization modes while in the
bottom panel the total flux is shown. Note that the fluxes shown are
in accordance with the definition of $H$ given in $\S 2.3$ and the
total radiative flux is given by $4 \pi H$; the subscript $E$ as usual
denotes the energy dependence.  The detailed features of the spectra
result mainly from the combined effects of the temperature profile,
the scattering between polarization modes, and the polarization of the
magnetic vacuum, which we discuss below. The magnetic field strength
enters all of these factors and thus affects the emerging spectra in
multiple ways.

Two results hold in all cases: {\it (i)} the extraordinary mode
carries most of the radiative flux and {\it (ii)} the spectra are not
Planckian, showing various amounts and shapes of hard excess. The
first result arises from the fact that the opacity of the
extraordinary mode is significantly smaller and hence it decouples
from the plasma at greater depths, where the temperature is higher
(Fig.~\ref{Fig:spec}). The second result is primarily, but not
exclusively, due to a similar reason: as shown previously (e.g.,
Shibanov et al.\ 1992; Rajagopal \& Romani 1997), the high-energy
photons originate deeper in the atmosphere due to their smaller
interaction cross section and hence the emerging spectra are bluer
than the blackbody spectrum at the effective temperature. The spectral
hardening because of the energy-dependence of the opacities is less
pronounced in the case of high magnetic fields than non-magnetic
neutron star atmospheres because the dependence of the magnetic
free-free opacity on photon energy $(K \propto E^{-1})$ is weaker than
that of the non-magnetic one $(K \propto E^{-3})$ (see also Zane,
Turolla, \& Treves 2000). However, it is important to note that in the
case of ultrastrong magnetic fields, different photospheres for
different energy photons is not the only origin of hard excess in the
spectra and the hard tails do not decrease monotonically with
increasing magnetic field strength. Instead, the spectral shape is
determined by two other effects.  First, as we will show in \S 4.3,
the broad-band absorption due to vacuum resonance changes the spectral
shape to reduce the color temperature $T_c$ of the best-fit blackbody
but gives rise to a pronounced power-law tail. This is because of the
energy- and density-dependence of the vacuum resonance as shown in
Figure~2. Second, scattering of photons from the extraordinary mode to
the ordinary mode, especially enhanced in the presence of vacuum
polarization resonance, reduces the spectral hardening. We also show
this below.

As the bottom panel in Figure~\ref{Fig:spec} shows, the effective
temperature controls the total emerging flux, as well as the peak and
broadness of the spectrum. Although the scaling of the total flux with
$T_{\rm eff}$ is straightforward, the flux at each photon energy does
not scale in the same way. This is because the temperature profiles
corresponding to different $T_{\rm eff}$ differ by more than a simple
displacement (see \S 4.1). Specifically, the outer layers can be
significantly hotter for higher effective temperatures, giving rise to
more flux at low photon energies. For a similar reason, the spectra
emerging from atmospheres with higher $T_{\rm eff}$ are more peaked,
with an increasing fraction of the total flux emerging around the peak
photon energy.

The effects of scattering between different polarization modes are
strongest at the lowest magnetic-field strengths, because, for a given
photon energy, the off-diagonal element of the scattering coefficient
from the extraordinary mode to the ordinary mode increases with
decreasing field strength. On the other hand, the effects of the
polarization of the magnetic vacuum are strongest at the highest field
strengths, as discussed in \S2.2. In order to disentangle these two
effects, we discuss the lowest ($10^{13}$~G) and the highest
($10^{15}$~G) magnetic field strengths in more detail, thus isolating
each effect.

\begin{figure}[t]
\centerline{ \psfig{file=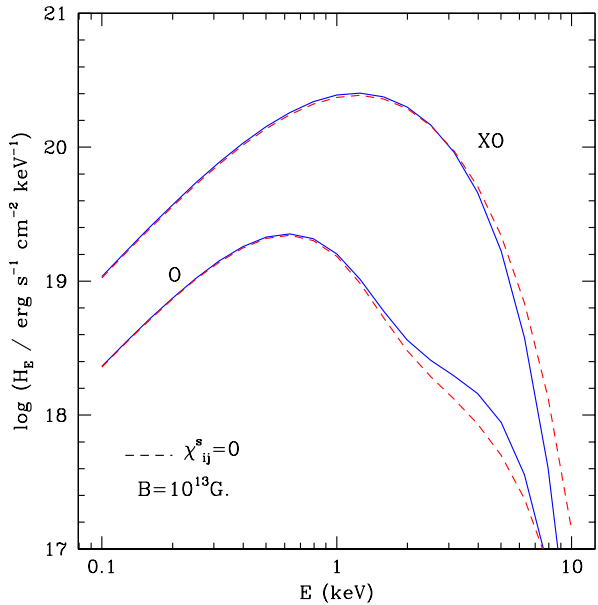,angle=0,width=8.5truecm} }
\figcaption[]{The spectra emerging from a model atmosphere with 
$T_{\rm eff}=0.3$~keV and $B=10^{13}$~G, when scattering between
different polarization modes is included (solid line) or neglected
(dashed line). \label{Fig:sscatt}}
\end{figure}

We use the temperature profile of the model atmosphere that
corresponds to $B=10^{13}$~G and $T_{\rm eff}=0.3$~keV and calculate
the emerging spectra, including or neglecting the scattering between
polarization modes. As can be seen in Figure~\ref{Fig:sscatt}, the
scattering between modes increases the flux of the ordinary mode at
high energies. This occurs because the off-diagonal elements of the
scattering coefficients increase, when $E$ approaches $E_{\rm b}$.  An
observable result of this cross-mode scattering is a small suppression
of the hard excess at $B \approx 10^{13}$~G. The discussion of the
vacuum polarization effects which are strong at $B \sim 10^{15}$~G are
presented in the next section.

\subsection{Effects of Vacuum Polarization on the Spectrum}

\begin{figure}[t]
\centerline{ \psfig{file=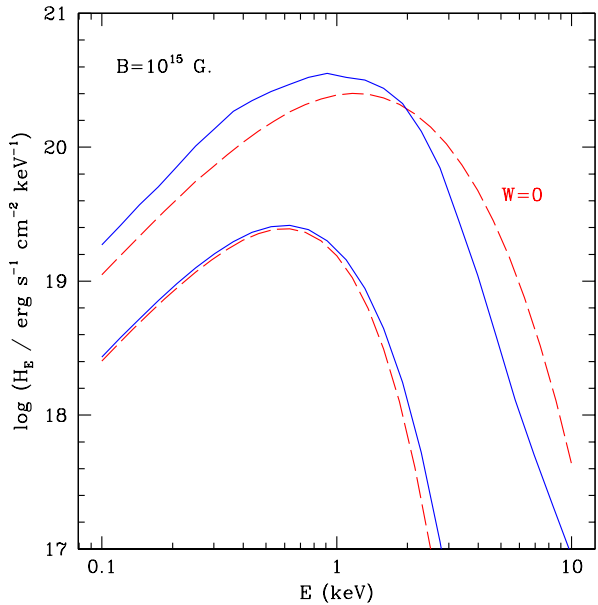,width=8.5truecm} }
\figcaption[]{The spectra emerging from a model atmosphere with 
$T_{\rm eff}=0.3$~keV and $B=10^{15}$~G, when the effects of vacuum
polarization are included (solid lines) or neglected (dashed
lines). \label{Fig:svac}}
\end{figure}

Taking into account the contribution of vacuum polarization to the
magnetic opacities leads to significant modification in the spectrum
of radiation emerging from the neutron star atmosphere. To demonstrate
the effects of vacuum resonance, we show in Figure~\ref{Fig:svac} two
model spectra where include (solid line) or neglect (dashed line) the
vacuum contribution to the scattering and absorption cross sections,
using model atmospheres that corresponds to $B=10^{15}$~G and $T_{\rm
eff}=0.3$~keV. The result is a drastic change in the overall spectral
shape: an increase of the flux at $ E \lesssim 2$~keV, an attenuation
of the flux at $E \gtrsim 2$~keV and the emergence of a hard spectral
tail at $E \gtrsim 3$~keV due to broadband absorption across the
vacuum resonance.

As Figure~2 shows, the main effect of the contribution of the vacuum
polarization to the magnetic opacities for most angles of photon
propagation is the emergence of a resonance at a photon-energy
dependent critical density in the plasma.  When the effects of vacuum
polarization are included, at each photon energy, the opacity of the
extraordinary mode becomes comparable to that of the ordinary mode and
the scattering from the extraordinary mode into the ordinary mode is
enhanced for a narrow range of densities. This rapid increase in
opacity of the extraordinary mode photons typically occurs at low
densities further out in the atmosphere (see Fig.~2), and thus at a
density where these photons have already decoupled from the
atmosphere. Therefore, the reduction in the flux arises from the fact
that going outward in the atmosphere, the essentially free-streaming
extraordinary mode photons encounter a resonance and either change
polarization mode or are absorbed over a narrow range of $\tau_T$.  In
the latter case, they decouple again outside the resonant layer and
stream outward. Note that in the absence of temperature adjustments in
the atmosphere, flux in the extraordinary mode at nearly all photon
energies would suffer such a reduction. On the other hand, at the
densities where the vacuum effects become important, the ordinary mode
is mostly optically thick and the reduction of the ordinary-mode
opacity due to vacuum contribution is minimal. As a result, the
primary effect of vacuum polarization is an attenuation of the flux
carried by the extraordinary mode at high photon energies.

The radiative equilibrium condition introduces an additional effect
and a further modification of the emerging spectrum.  This is because
the atmosphere adjusts to the increase in opacity, and the
corresponding reduction in the flux, by an overall increase in the
temperature in the resonant layers (see \S 4.1 and
Fig.~\ref{Fig:tvac}). This can also be understood as a heating of the
atmosphere due to resonant absorption.  This in turn leads to an
increase of the flux carried by the low-energy photons (so that the
energy-integrated flux is constant) and therefore to an overall shift
of the spectrum towards lower energies and a further distortion away
from a Planckian spectrum.

\begin{figure}[t]
\centerline{ \psfig{file=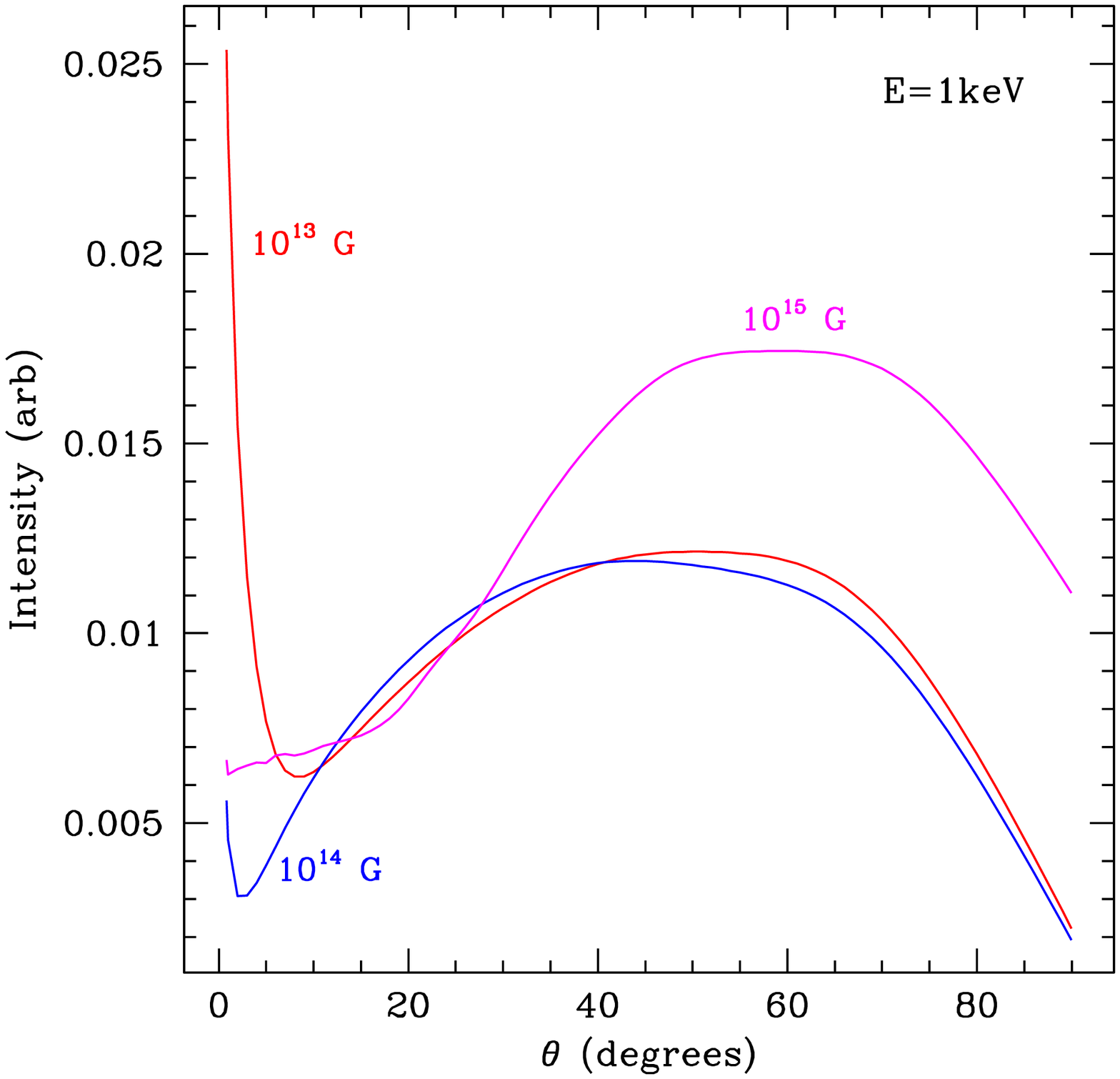,width=8.0truecm} }
\centerline{ \psfig{file=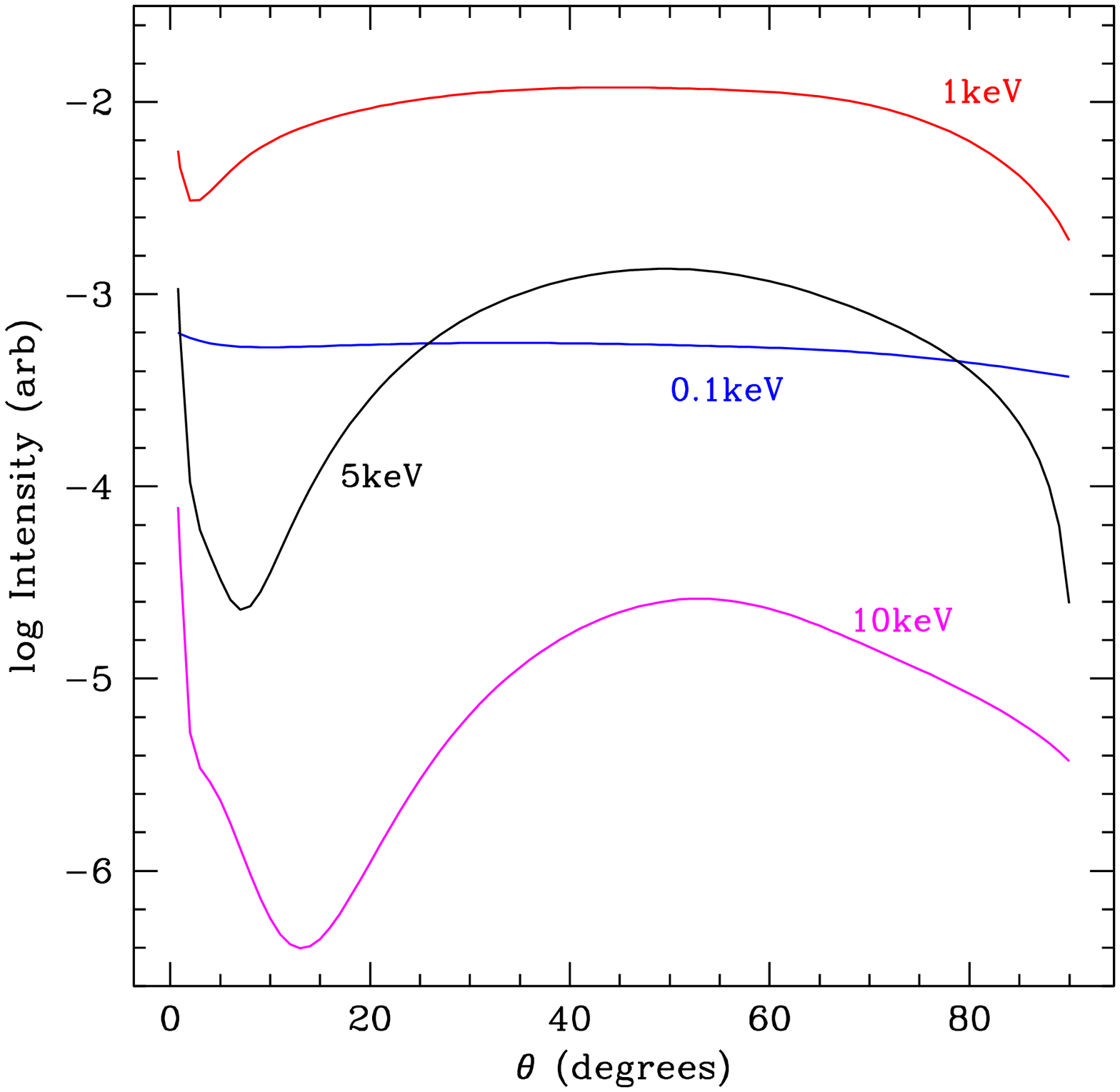,width=8.0truecm} }
\figcaption[]{(top panel) Angle dependence of emerging intensity for 
$T_{\rm eff} = 0.3$ keV at $B=10^{13}$G (short-dashed line),
$B=10^{14}$ G (long-dashed line), $B=10^{15}$ G (solid line). (bottom
panel) Same as above, with $B=10^{14}$ G and $E=0.1$~keV, 1~keV, 5~keV
and 10~keV. \label{Fig:beam}}
\end{figure}

\subsection{The Beaming of Emerging Radiation}

Finally, we discuss the angular pattern of the radiation emerging from
the different model atmospheres. In Figure~\ref{Fig:beam} (top panel),
we show the beaming of radiation at a photon energy of 1~keV and for
three different magnetic-field strengths, while in the bottom panel,
we show the beaming of radiation emerging from a model atmosphere with
$B=10^{14}$~G, at three different photon energies. Note that the
emerging intensity in the ordinary and extraordinary modes are added
together in these figures.

Both figures show that the radiation pattern consists of two
components: a narrow pencil beam at small angles ($\lesssim 5^\circ$)
and a broader fan beam at intermediate angles ($\sim 20-60^\circ$).
This pattern is a direct consequence of the angular dependence of the
cross sections in the two modes (see Fig.~\ref{Fig:opac}), combined
with the temperature profiles in the atmospheres
(Fig.~\ref{Fig:temp}). In particular, the extraordinary mode
contributes to both the pencil and the fan beams, because its opacity
is peaked at an angle away from $\theta=0$, decreasing towards both
smaller and larger angles. On the other hand, the ordinary mode
contributes (roughly equally as the extraordinary mode) to the narrow
pencil beam of the pattern, because its opacity decreases
significantly at small angles. Therefore while not contributing
significantly to the total flux (see \S4.2), the ordinary mode plays
an important role in determining the beaming pattern with its large
specific intensity near $\theta=0$.

The angular dependence of the magnetic cross sections, and hence the
beaming of radiation, depend both on photon energy and on magnetic
field strength, through the ratio $E/E_{\rm b}$. As
Figure~\ref{Fig:beam} shows, at photon energies much smaller than the
electron cyclotron energy, the pencil beam is constrained to
increasingly smaller angles, $\sin\theta \leq (E/E_{\rm b})^{1/2}$
while the fan beam becomes more prominent, resulting in a flatter
radiation pattern. Therefore, for a given magnetic field, the emerging
radiation is more strongly and radially beamed at high photon
energies, and for a given photon energy the radiation is more beamed
at lower magnetic field strengths. Note that with increasing field
strength or decreasing photon energy, the peak value of the pencil
beam may in fact increase; however the total flux in the pencil beam
decreases, thus making it less prominent. Numerically, the pencil beam
in such cases, constrained to $\theta \lesssim 0.1^\circ$, becomes
hard to resolve and has minimal observational consequences.

In addition, at $B \simeq 10^{15}$~G, vacuum polarization affects the
beam shape at photon energies $E \gtrsim 1$~keV. Specifically, the
scattering cross sections are enhanced due to the vacuum contribution,
redistributing the photons between the two polarization modes and
different angles of propagation more efficiently. This results in a
beam shape with a broader pencil component and a broader plateau
between the pencil and fan beams than that of the $B=10^{14}$~G case
(Fig.~8) and the case when the vacuum contribution is ignored.

\begin{figure}[t]
\centerline{ \psfig{file=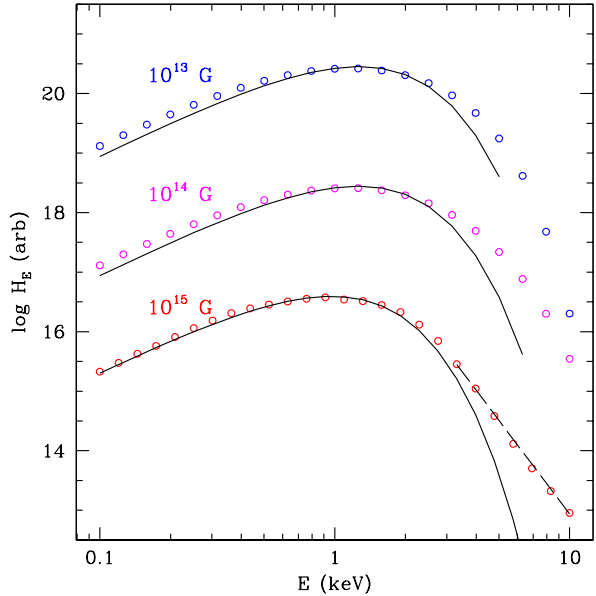,width=8.5truecm} }
\figcaption[]{ \footnotesize The spectra (open circles) emerging 
from \ns atmospheres with $T_{\rm eff} = 0.3$~keV and different
magnetic field strengths, along with the best-fit black body spectra
in the photon energy range $0.5-10$~keV (solid lines). At $10^{15}$~G,
the dashed lines represent the best-fit power-law spectral tail at
high photon energies. All spectra are displaced vertically for
clarity. \label{Fig:sfit}}
\end{figure}

\section{Discussion and Conclusions}

We have constructed model atmospheres for strongly magnetic \nss\ in
radiative equilibrium. We have considered the complete angle,
polarization and energy dependence of the magnetic free-free
absorption and scattering processes, taking into account the effects
of polarizability of the magnetic vacuum. We find that the resulting
spectra are broader and bluer than the black body spectrum at the
effective temperature, and depend on the magnetic field strength.

Figure~\ref{Fig:sfit} shows the dependence of the emerging spectra on
magnetic field strength for $T_{\rm eff} = 0.3$~keV along with the
best-fit black body spectra in the photon energy range $0.5-10$~keV,
displaced vertically for clarity. In determining the best fit
blackbody and the color temperature $T_{\rm c}$, we fit the peak of
the spectra and allow for a hard excess, either in the form of a
power-law tail or a broad excess. For $B \lesssim 10^{14}$~G, the
spectra have distorted Planckian shapes and are generally broader than
a black body. Therefore, when fitted with black body shapes, the
best-fit (color) temperatures, $T_{\rm c}$ depend on the range of
photon energies used (see Table~1). Typically, for these magnetic
fields, $T_{\rm c}/T_{\rm eff} \simeq 1.5-1.8$.

At $B > 10^{14}$~G, the spectra are significantly modified by
the vacuum polarization effects and are in general well-described by a
black-body shape at low energies with a high energy power-law tail.
The corresponding color temperatures are smaller (see Table~1), with
$T_{\rm c}/T_{\rm eff} \simeq 1.2$ at $B=10^{15}$~G. For this magnetic
field strength, the power-law tail has an energy index of $-4$.

At all magnetic field strengths, the angular dependence has two
maxima: a narrow (pencil) beam at small angles ($\lesssim 5^\circ$)
with respect to the normal and a broad maximum (fan beam) at
intermediate angles ($\sim 20-60^\circ$). The relative importance and
the opening angle of the radial beam decreases strongly with
increasing magnetic field strength and decreasing photon energy (see
\S 4.3). As a result, the emerging radiation from a
thermally-emitting neutron star has a prominent radial peak only at
relatively low magnetic field strengths ($B \sim 10^{13}$~G) and at
high photon energies ($E \gtrsim 1$~keV) within the range of $B$ and
$E$ values considered here.

Our results have significant implications for models of thermally
emitting neutron stars with strong magnetic fields, and specifically
for magnetar models of SGRs and AXPs. First, the observed power-law
spectral tails at high photon energies need not arise from a
non-thermal (possibly magnetospheric) emission mechanism but may be
the result of vacuum polarization at very strong magnetic fields (see
Fig.~9). Second, the beaming of the emerging radiation determines the
pulse shapes and amplitudes of spinning neutron stars with anisotropic
surface temperature distributions. These properties depend also on the
bending of photon trajectories in the gravitational field of neutron
stars and can be used in constraining models of SGRs and AXPs (DeDeo,
Psaltis, \& Narayan 2001; Psaltis, \"Ozel, \& DeDeo 2000). We will
report the results of a detailed comparison of the observed spectral
and timing properties of SGRs and AXPs with the predictions of the
models discussed here in a forthcoming paper (\"Ozel, Psaltis, \&
Kaspi 2001).

\acknowledgements
I am grateful to Ramesh Narayan for countless stimulating discussions
and especially for his help in developing the temperature correction
scheme.  I am also grateful to Dimitrios Psaltis for his help in many
aspects of constructing and understanding models of neutron star
atmospheres throughout this work. I thank George Rybicki and Dimitar
Sasselov for useful discussions on radiative transfer. I have
benefited a lot from discussions with George Pavlov, Dong Lai, and
Roger Romani on magnetic processes during the conference on ``Spin and
Magnetism in Young Neutron Stars'' at ITP. I also thank Vicky Kaspi,
Cole Miller, George Pavlov, and Peter Woods for useful comments on the
manuscript. This work was supported in part by NSF Grant AST 9820686.

\newpage
\begin{deluxetable}{cccc}
\tablecaption{Color Temperatures\tablenotemark{a}}
\tablehead{$T_{\rm eff}$ & $T_{\rm c}/T_{\rm eff}(10^{13}$~G) 
& $T_{\rm c}/T_{\rm eff}(10^{14}$~G) 
& $T_{\rm c}/T_{\rm eff}(10^{15}$~G) }
\startdata
0.1~keV & 1.5/1.8 & 1.5/1.8 & 1.4/1.7\\
0.3~keV & 1.4/1.5 & 1.5/1.5 & 1.1/1.1\\
0.5~keV & 1.4/1.4 & 1.6/1.5 & 1.1/1.1
\enddata

\tablenotetext{a}{The two color temperatures in each entry 
corresponds to the best fit blackbody spectra in the ranges 
0.1--10~keV and 0.5--10~keV, respectively.}

\label{Tcolor}
\end{deluxetable}

\end{document}